\documentclass[prl,reprint,a4paper]{revtex4-1}
\usepackage{amsmath}   \usepackage{amsthm}    \usepackage{amssymb}   \usepackage{graphicx}
\usepackage{psfrag}  \usepackage{color}
\usepackage{tensor}
\usepackage{listings}
\usepackage{mathrsfs}
\usepackage{amsmath}
\usepackage{calc}
\usepackage{keyval}
\usepackage{xcolor,color}
\usepackage{natbib}
\usepackage[]{subfig}
\usepackage[breaklinks=true]{hyperref}
    \hypersetup{
        unicode=false,                  pdftoolbar=true,                pdfmenubar=true,                pdffitwindow=false,             pdfstartview={FitH},            pdfcreator={Creator},           pdfkeywords={keyword1, key2, key3},         pdfnewwindow=true,              colorlinks=true,                linkcolor=black,              citecolor=black,              filecolor=magenta,              urlcolor=cyan               }
\usepackage{cleveref}
    \crefname{section}{Section}{Sections}
    \Crefname{section}{Section}{Sections}
    \crefname{chapter}{Chapter}{Chapters}
    \Crefname{chapter}{Chapter}{Chapters}
    \crefname{figure}{Fig.}{Figs.}
    \Crefname{figure}{Figure}{Figures}
    \crefname{equation}{Eq.}{Eqs.}
    \Crefname{equation}{Equation}{Equations}
    \crefname{table}{Tab.}{Tabs.}
    \Crefname{table}{Table}{Tables}
    \setcitestyle{citesep={,}}

\usepackage{tikz}
\usetikzlibrary{decorations.pathreplacing}

\definecolor{myorange}{rgb}{0.93, 0.69, 0.13}
\definecolor{mygrey}{rgb}{0.5, 0.5, 0.5}

 \renewcommand{\v}[1]{\ensuremath{\mathbf{#1}}} \newcommand{\gv}[1]{\ensuremath{\boldsymbol{#1}}} 
\newcommand{\uv}[1]{\ensuremath{\mathbf{\hat{#1}}}}

 \newcommand{\pd}[2]{\frac{\partial #1}{\partial #2}}

\newcommand{\dinline}[2]{{\mbox{d}{#1}}/{\mbox{d}{#2}}}

\newcommand\solidrule[1][0.3cm]{\rule[0.5ex]{#1}{.8pt}}
\newcommand{\blackline} {{\color{black}\solidrule}}
\newcommand{\redline} {{\color{red}\solidrule}}

\newcommand{\blueline} {{\color{blue}\solidrule}}
\newcommand{\orangeline} {{\color{myorange}\solidrule}}
\newcommand{\greyline} {{\color{mygrey}\solidrule}}
\newcommand\dashedrule{\mbox{\solidrule[1mm]\hspace{.5mm}\solidrule[1mm]\hspace{.5mm}\solidrule[1mm]}}

\newcommand{\blackdashedline} {{\color{black}\dashedrule}}

\newcommand\dashedtworule{\mbox{\solidrule[2mm]\hspace{.5mm}\solidrule[1mm]\hspace{.5mm}\solidrule[2mm]}}

\newcommand{\blackdashedtwoline} {{\color{black}\dashedtworule}}
\newcommand{\greydashedtwoline} {{\color{mygrey}\dashedtworule}}
\newcommand\dashedthreerule{\mbox{\solidrule[.3mm]\hspace{.25mm}\solidrule[.3mm]\hspace{.25mm}\solidrule[.3mm]\hspace{.25mm}\solidrule[.3mm]\hspace{.25mm}\solidrule[.3mm]\hspace{.25mm}\solidrule[.3mm]}}
\newcommand{\blackdashedthreeline} {{\color{black}\dashedthreerule}}

\newcommand{\opencircle} {{\color{black}$\boldsymbol\circ$}}

\newcommand{\filledbluesquare} {{\color{blue}$\blacksquare$}}
\newcommand{\filledblacksquare} {$\blacksquare$}
\newcommand{\filledredsquare} {{\color{red}$\blacksquare$}}
\newcommand{\filledorangesquare} {{\color{myorange}$\blacksquare$}}
\newcommand{\filledgreysquare} {{\color{mygrey}$\blacksquare$}}
\newcommand{\opensquare} {{\color{black}$\square$}}

\newcommand{\opengreytriangleup} {{\color{mygrey}$\triangle$}}

\newcommand{\greycross} {{\color{mygrey}\textbf{+}}}

\newcommand{\bluestar} {{\color{blue}$*$}}

\newcommand{\mypara}{{\mkern3mu\protect\vphantom
                     {\perp}\vrule depth 0pt\mkern2mu\vrule depth 0pt\mkern3mu}}

\newcommand{\vort}{\gv{\omega}}
\newcommand{\dissip}{\varepsilon}
\newcommand{\dissipinv}{\dissip_{inv}}
\newcommand{\dissipnu}{\dissip_{\nu}}
\newcommand{\dissipI}{\dissip_{I}}

\newcommand{\wn}{\kappa}

\newcommand{\wnpp}{\wn_\perp}
\newcommand{\wnpar}{\wn_\mypara}

\newcommand{\wndpp}{\wn_f \mathcal{L}_\perp}
\newcommand{\wndpar}{\wn_f \mathcal{L}_\mypara}

\newcommand{\kzeman}{\wn_{\Omega}}
\newcommand{\wnf}{\wn_f}

\newcommand{\lf}{\ell_f}
\newcommand{\tf}{\tau_f}

\newcommand{\lkolmo}{\eta}
\newcommand{\tomg}{\tau_\Omega}
\newcommand{\twave}{\tau_w}
\newcommand{\uf}{u_f}

\newcommand{\uell}{u_\ell}

\newcommand{\romanreynoldseps}{\mathrm{Re_{\varepsilon}}}
\newcommand{\romanreynoldslmbda}{\mathrm{Re_{\lambda}}}
\newcommand{\romanrossbyeps}{\mathrm{Ro_{\varepsilon}}}
\newcommand{\romanrossbylmbda}{\mathrm{Ro_{\lambda}}}

\usepackage{soul}
\usepackage{xcolor}

\begin{document}

\title{\textbf{Regime Transition in the Energy Cascade of Rotating Turbulence}}
\author{T. Pestana}
\author{S. Hickel}
\affiliation{Aerodynamics Group, Faculty of Aerospace Engineering,
Delft University of Technology,
Kluyverweg 2, 2629 HS Delft, The Netherlands}
\date{\today}

\begin{abstract}
Transition from a split to a forward kinetic energy cascade system is explored
in the context of rotating turbulence using direct numerical simulations with a
three-dimensional isotropic random force uncorrelated with the velocity field.
Our parametric study covers confinement effects in large aspect ratio domains
and a broad range of rotation rates. The data here presented add substantially
to previous works, which, in contrast, focused on smaller and shallower domains.
Results indicate that for fixed geometrical dimensions the Rossby number acts as
a control parameter, whereas for a fixed Rossby number the product of the domain
size along the rotation axis and forcing wavenumber governs the amount of energy
that cascades inversely. The regime transition criterion hence depends on both
control parameters.
\end{abstract}

\maketitle

\section{Introduction}

The energy cascade is the fundamental mechanism in turbulent flows that
describes the energy exchange between the various scales of motion
\citep{Frisch1995}. A forward cascade from large to small scales is commonly
observed in three-dimensional (3D) flows, whereas an inverse energy cascade from
small towards large scales is the hallmark of two-dimensional (2D) flows
\citep{Alexakis2018,Boffetta2012}. Predicting the energy cascade direction,
therefore, requires anticipating if, for a given set of control parameters, the
resulting flow field resembles best 3D or 2D flow dynamics. In lack of
analytical predictions, a typical approach consists of carefully designing
numerical experiments, where the system's parameters are individually varied to
produce a phase transition diagram. Throughout this study we consider a large
number of forced direct numerical simulations (DNS) and analyze the influence of
geometric confinement and system rotation on the cascade direction in
homogeneous rotating turbulence.

Inertial waves, i.e. plane wave solutions to the linearized Navier-Stokes
equations, can modulate the energy transfer in rotating turbulence
\citep{greenspan1968theory,Godeferd2015}. By considering high rotation rates and
exploiting the fact that rotating turbulence is a multi-timescale problem,
\citet{Waleffe1993} suggested that the nonlinear dynamics are modified by wave
interactions. Resonant wave interactions can explain the favored energy transfer
towards horizontal modes, whereas non-resonant wave interactions are considered
to damp and inhibit the triadic interactions typical of homogeneous turbulence
\citep{Cambon1997,Smith1999}. This mechanism also persists at lower rotation
rates due to homochirical interactions that transfer energy into the plane
orthogonal to the rotation axis \citep{Buzzicotti2018}. As a consequence, when
rotating homogeneous flows are forced at wavenumber $\wnf$, the injected energy
can cascade both to larger ($\wn<\wnf$) and smaller scales
($\wn>\wnf$); this is hereafter referred to as split energy cascade. These
findings help to explain the preferential upscale of energy typically
found in numerical and experimental investigations of rotating turbulent flows
\citep{Yeung1998,Smith1999,Mininni2009,Moisy2011,Mininni2012a,Delache2014}.
Nevertheless, we must bear in mind that a large network of
triadic interactions as in the Navier-Stokes equations can evolve differently
than a set of isolated triads, as previously pointed out in Refs.
\citep{Linkmann2017,Moffatt2014}.

Among different theories that elucidate the phenomenon of rotating turbulence,
the work of \citet{Galtier2003} is regarded as an important contribution. Based
on wave turbulence theory, which deals with systems where interactions are
governed by waves, he derived scaling laws for the energy spectrum. These laws
were also shown to follow from phenomenological arguments for the spectral
transfer time~\textemdash~a typical energy transfer timescale. For infinitely
large domains, as required by wave turbulence theory
\citep{waveturb_nazarenko2011}, the weak inertial-wave theory of
\citet{Galtier2003} predicts that energy cascades forward and to small scales.
However, a passage from a split to a forward energy cascade system by approaching
the large-box limit has not yet been confirmed by DNS.

In the absence of rotation, however, the geometrical dimensions of the system
itself influences the energy cascade direction. Using a two-dimensional
two-component (2D2C) horizontal force, \citet{Smith1996a} and \citet{Celani2010}
found that the ratio $L_3/\lf$, where $L_3$ is the vertical domain extension and
$\lf$ is the forcing lengthscale, is a governing control parameter. They showed
that large $L_3/\lf$ results in a forward energy cascade, whereas inverse energy
transfer was triggered and split the energy cascade for $L_3/\lf \leq 1/2$. More
recently, numerical simulations by \citet{Benavides2017} explored transitions in
a thin layer of fluid subjected to free-slip boundary conditions. Transition
from a forward to a split energy cascade was shown to be critical and depend on
the ratio of forcing lengthscale to wall separation.

Regime transitions in rotating homogeneous turbulence are therefore affected by
geometrical dimensions and rotation rate. \citet{Deusebio2014} studied
hyperviscous fluids in rotating small aspect ratio domains subjected to 2D2C
forcing and found that large rotation rates as well as small $L_3/\lf$ suppress
enstrophy production and induce an inverse energy cascade. Their data proves, at
least for weak rotation rates, that transition from a split to a forward cascade
is possible by controlling either rotation rate or domain size. For strong
rotation, however, almost the entire injected energy cascaded inversely.
Although transition was not observed, they hypothesized that it could still take
place for sufficiently large $L_3/\lf$. This conjecture, however, remains to be
verified by either forcing smaller flow scales or by increasing the domain size
\citep{Seshasayanan2018}.

The present work sheds light on the question whether a transition from a split
to a forward cascade system always exists in forced homogeneous rotating
turbulence. We conduct a systematic parametric study that covers several
rotation rates and an unprecedented range of geometric confinements by
considering strongly elongated domains and large forcing wavenumbers $\wnf$.
This new database is complementary to previous studies, which focused on the
confinement induced transition in smaller and shallower domains. Through
large-scale forcing, we construct isotropic flow fields that are posteriorly
subjected to rotation. Differently from previous studies, we employ a
three-dimensional three-component (3D3C) forcing scheme that by design provides
a constant energy input independent of the velocity field. We believe this
results in a neater and more general framework where anisotropy originates
solely from rotation. 
\section{Methodology and Governing Parameters}
We solve the incompressible Navier-Stokes equations in a frame rotating at rate
$\gv{\Omega}$:
\begin{align}
     \nabla \cdot \v{u} &= 0, \label{eq:mass} \\
     \pd{\v{u}}{t} + (2\,\gv{\Omega} + \vort) \times \v{u} &= - \nabla q
     + \nu \nabla^2 \v{u} + \v{f}.
     \label{eq:ns}
\end{align}
Here, $\v{u}$, $\vort$ and $\v{f}$ are velocity, vorticity and an external
force, respectively. The reduced pressure into which the centrifugal force is
incorporated is given by $q$, and $\nu$ denotes the kinematic viscosity.

\Cref{eq:mass,eq:ns} are discretized in space by a dealiased Fourier
pseudo-spectral method (2/3-rule) in a triply-periodic domain of size
$2\pi\mathcal{L}_1 \times 2\pi\mathcal{L}_2
\times 2\pi\mathcal{L}_3$ \citep{Orszag1971,p3dfft}. The rotation axis is
assumed aligned with the vertical direction, i.e. $\gv{\Omega}=\Omega
\,\uv{e}_3$, and we restrict ourselves to cases where the domain size in the
direction perpendicular to the axis of rotation are equal:
$\mathcal{L}_1=\mathcal{L}_2=\mathcal{L}_\perp =1$. Accordingly,
$\mathcal{L}_\mypara$ replaces $\mathcal{L}_3$ to denote the domain size in the
direction parallel to the rotation axis, and can be arbitrarily
chosen. We use Rogallo's integrating factor technique for exact time
integration of the viscous and Coriolis terms and a third-order Runge-Kutta
scheme for the nonlinear terms \citep{Rogallo1977,Morinishi2001a}.

The external force $\v{f}$ injects energy to the system at rate $\dissipI$, see
Ref.~\citep{Alvelius1999}. The force's spectrum $F(\wn)$, from which $\v{f}$ in
\cref{eq:ns} is assembled, is Gaussian distributed, centered around a wavenumber
$\wnf$ and has standard deviation $c=0.5$: $F(\wn) = A \exp(-(\wn -
\wnf)^2/c)$.
\begin{table}[b]
\caption{List of direct numerical simulations at $\romanreynoldseps\approx55$.
The $\romanrossbyeps$ numbers are given in the footnote.}
\begin{ruledtabular}
\begin{tabular}{lccccc}
Case & $\wnf\mathcal{L}_\perp$  & $\wnf\mathcal{L}_\mypara$ & $A_r$ &  $N_p$
\tabularnewline
\hline\\[-2ex]
\texttt{kf02-a01} \footnote{$\romanrossbyeps\approx0.31,\, 0.06$
                  \label{fn:weak}} & 2 & 2 &   1  & $192^3$  \tabularnewline
\\[-2ex]
\texttt{kf04-a01} \footnotemark[1]
                  & 4  & 4 & 1   &  $384^3$                 \tabularnewline

\texttt{kf04-a02} \footnote{$\romanrossbyeps\approx0.06$ \label{fn:strong}}
                  & 4  & 8 & 2  &  $384^2  \times 768$     \tabularnewline

\texttt{kf04-a04} \footnotemark[2]
                  & 4  & 16 & 4  &  $384^2  \times 1536$    \tabularnewline

\texttt{kf04-a08} \footnotemark[2]
                  & 4  & 32 & 8  &  $384^2  \times 3072$    \tabularnewline

\texttt{kf04-a16} \footnotemark[2]
                  & 4  & 64 & 16  &  $384^2  \times 6144$    \tabularnewline

\texttt{kf04-a32} \footnotemark[2]
                  & 4  & 128 & 32 &  $384^2  \times 12288$   \tabularnewline
\\[-2ex]

\texttt{kf08-a01} $^a$                   & 8  & 8 & 1  & $768^3$                 \tabularnewline

\texttt{kf08-a02} \footnotemark[2]
                  & 8  & 16 & 2 & $768^2 \times  1536$    \tabularnewline

\texttt{kf08-a04} \footnotemark[2]
                  & 8  & 32 & 4 & $768^2 \times  3072$    \tabularnewline

\texttt{kf08-a08} \footnote{$\romanrossbyeps \approx
                  1.25,\, 0.63,\, 0.31,\,
                  0.27,\, 0.24,\, 0.22,\, 0.19,\, 0.16,\, 0.14,\, 0.11,\,$
                  $0.09,\, 0.08,\, 0.06$}
                  & 8  & 64 & 8 & $768^2 \times  6144$    \tabularnewline

\texttt{kf08-a16} \footnotemark[2]
                  & 8  & 128 & 16 & $768^2 \times  12288$   \tabularnewline
\\[-2ex]

\texttt{kf16-a01} \footnotemark[1]
                  & 16 & 16 & 1 & $1536^3$                \tabularnewline

\texttt{kf16-a02} \footnotemark[2]
                  & 16 & 32 & 2 & $1536^2 \times  3072$   \tabularnewline

\texttt{kf16-a04} \footnotemark[2]
                  & 16 & 64 & 4 & $1536^2 \times  6144$   \tabularnewline
\\[-2ex]
\texttt{kf32-a01} \footnotemark[2]
                  & 32 & 32 & 1 & $3072^3$   \tabularnewline

\end{tabular}
\end{ruledtabular}

 \label{tb:perfosimul}
\end{table}
For given $\wnf$ and $c$, the prefactor $A$ is uniquely determined from the
desired energy input rate $\dissipI$. In the absence of rotation, we obtain
isotropic velocity fields and a balance between energy input rate and viscous
dissipation, i.e. $\dissipI=\dissipnu$. This forcing scheme ensures through
projection that the force and velocity field are uncorrelated at every instant
of time \citep{Alvelius1999}. As a consequence, $\dissipI$ is solely determined
by the force-force correlation and is independent of the velocity
field. Thus, we can define a priori true control parameters from which the
governing non-dimensional numbers are derived.

The domain size, $\mathcal{L}_\mypara$ and $\mathcal{L}_\perp$, the forcing
wavenumber $\wnf$, the viscosity $\nu$, the rotation rate $\Omega$ and the
energy input rate $\dissipI$ can all be freely chosen. Regarding $\dissipI$,
it could be additionally decomposed in three contributions stemming from the
power injected in each direction. However, because the forcing is isotropic, it
is sufficient to consider the total power input $\dissipI$ only. These six
parameters $\{\wnf,\nu,\dissipI,\Omega,\mathcal{L}_\perp,\mathcal{L}_\mypara\}$
form the set of true control parameters and are the basis for the
non-dimensional similarity numbers. The characteristic length, velocity and
time-scale follow naturally as $\lf = \wnf^{-1}$, $\uf =
\dissipI^{1/3} \,\, \wnf^{-1/3}$, and $\tf =
\wnf^{-2/3}\,\,\dissipI^{-1/3}$, respectively. In addition, a timescale based
on the rotation rate is taken as $\tomg = 1/(2\Omega)$.

The Reynolds and Rossby numbers are now unambiguously defined as
\begin{equation}
    \romanreynoldseps = \frac{\dissipI^{1/3} \,\, \wnf^{-4/3}}{\nu}
    \quad \mbox{and} \quad
    \romanrossbyeps = \frac{\wnf^{2/3}\dissipI^{1/3}}{2\Omega}.
\end{equation}
From the problem's geometry and the forcing wavenumber, we define two other
non-dimensional numbers, i.e. $\wn_{f}\mathcal{L}_{\perp}$ and
$\wn_{f}\mathcal{L}_\mypara$. Hence, we obtain a set of four independent
governing non-dimensional numbers that fully describes our numerical
experiments: $\romanreynoldseps$, $\romanrossbyeps$, $\wn_{f}\mathcal{L}_\perp$
and $\wn_{f}\mathcal{L}_\mypara$. As the final goal is to investigate
dimensional and rotational effects on forced homogeneous rotating turbulence, we
fix $\romanreynoldseps$ and allow $\romanrossbyeps$, $\wndpar$ and $\wndpp$ to
vary. We remark that this set is not unique and other non-dimensional groups
exist. For instance, $\romanreynoldseps$ and $\romanrossbylmbda$ could be
combined to form the micro-scale Rossby number
$\romanrossbylmbda=\romanreynoldseps^{1/2} \romanrossbylmbda$ (ratio of rotation
and Kolmogorov timescale \citep{Cambon1997}) or $\wndpar$ and $\wndpp$ could be
related to obtain the domain's aspect ratio $A_r
=
\mathcal{L}_\mypara/\mathcal{L}_\perp$. 
Initial conditions were generated by performing DNS of non-rotating forced
isotropic turbulence. We started from a zero-velocity field and marched in time
until a fully developed steady-state was achieved. After the initial transient
statistics, were sampled over at least $24\,\tf$, corresponding to
approximately ten large-eddy turnover times. Following this procedure, a
reference isotropic solution was computed for every entry in
\cref{tb:perfosimul}.

The initially imposed $\romanreynoldseps\approx55$ ultimately led to homogeneous
non-rotating turbulent fields with a characteristic Taylor micro-scale Reynolds
number $\romanreynoldslmbda\approx68$. The spatial resolution in terms of the
Kolmogorov lengthscale $\eta$ was kept constant throughout this study, i.e.
$\wn_{max}\eta\approx1.5$, where $\wn_{max}$ is the largest represented
wavenumber. For the case with largest $\wndpar$, the integral lengthscale in the
direction of rotation is about $600$ times smaller than the respective domain
size.

\begin{figure}[t]
  \centering
  \includegraphics{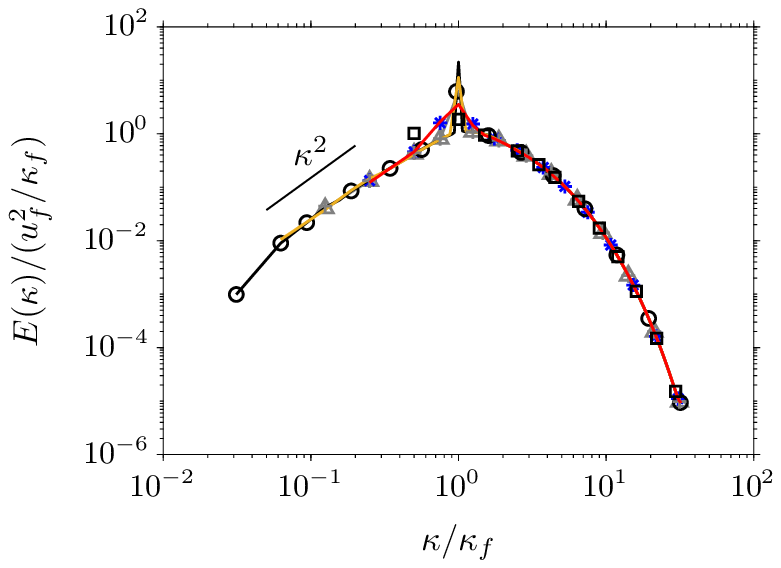}
  \caption{Three-dimensional spherically averaged energy spectrum of the
  initial condition:
  \texttt{kf02-a01} (\opensquare),
  \texttt{kf04-a01} (\redline),
  \texttt{kf08-a01} (\greycross),
  \texttt{kf16-a01} (\orangeline),
  \texttt{kf32-a01} (\blackline \opencircle \blackline),
  \texttt{kf04-a32} (\bluestar)
  \texttt{kf08-a16} (\opengreytriangleup)
   }
  \label{fig:iso_energyspec}
\end{figure}

\Cref{fig:iso_energyspec} compares the 3D spherically averaged energy spectrum
$E(\wn)$ for cases with aspect ratio $A_r=1$, which contain ``\texttt{a01}'' in
its name description, and two additional simulations with $A_r=16$ and $A_r=32 $
(cases \texttt{kf04-a32} and \texttt{kf08-a16} in \cref{tb:perfosimul}). This
data proves the equivalence between initial conditions for DNS forced at
different wavenumbers and those computed with distinct $\wndpar$ and $\wndpp$.
We find that the energy spectra perfectly coincide and that $E(\wn)$ scales best
with $\wn^2$ at wavenumbers $\wn<\wnf$, in agreement with
Ref.~\citep{Dallas2015}.
The obtained isotropic velocity fields were used as initial condition for the
simulations with different rotation rates.
The statistical variability of the results for small domains was reduced
by ensemble averaging. For the smallest domain \texttt{kf02-a01} we ensemble
averaged $10$ independent realizations and cases \texttt{kf04} with $A_r>1$ are
averages of $3$ realizations. For all other cases, the data represents a single
numerical experiment.
\begin{figure*}[t]
  \centering
  \captionsetup[subfigure]{labelformat=empty}
  \subfloat
          {\includegraphics{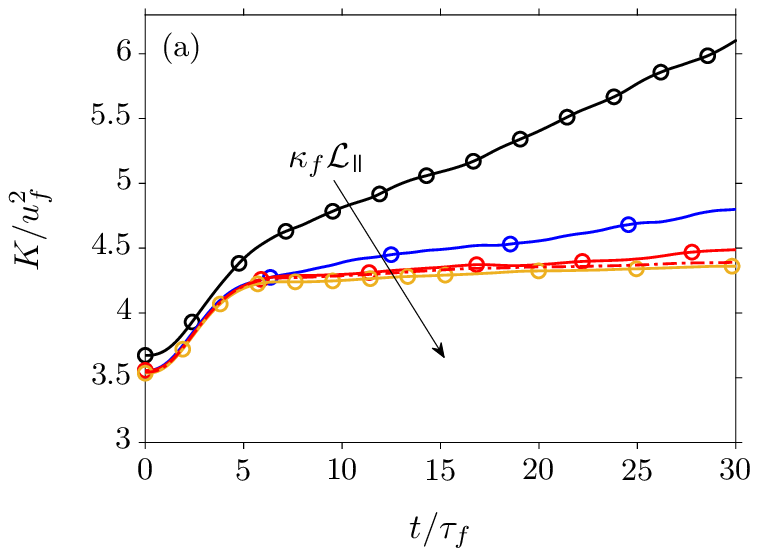}
          \label{fig:weak_kinetic_energy}}
    \hfill
  \subfloat
          {\includegraphics{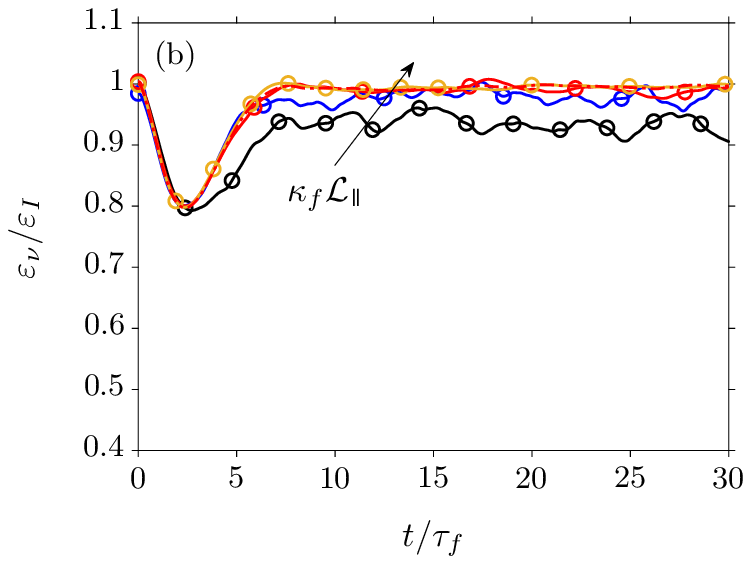}
          \label{fig:weak_dissipation}}
  \caption{Time evolution of box-averaged kinetic energy (a) and energy
  dissipation rate (b) for $\romanrossbyeps\approx0.31$ (weak rotation).
  Lines corresponding to same $\wndpp$ are grouped by color:
  $\wndpp=2$ (\filledblacksquare),
  $\wndpp=4$ (\filledbluesquare),
  $\wndpp=8$ (\filledredsquare),
  $\wndpp=16$ (\filledorangesquare),
  Lines corresponding to the same $A_r$ are grouped by line types:
  $A_r=1$ (\blackline\opencircle\blackline),
  $A_r=8$ (\blackdashedtwoline), cf. \cref{tb:perfosimul}.}
  \label{fig:weak}
\end{figure*}
\begin{figure*}[t]
  \centering
  \captionsetup[subfigure]{labelformat=empty}
  \subfloat
          {\includegraphics{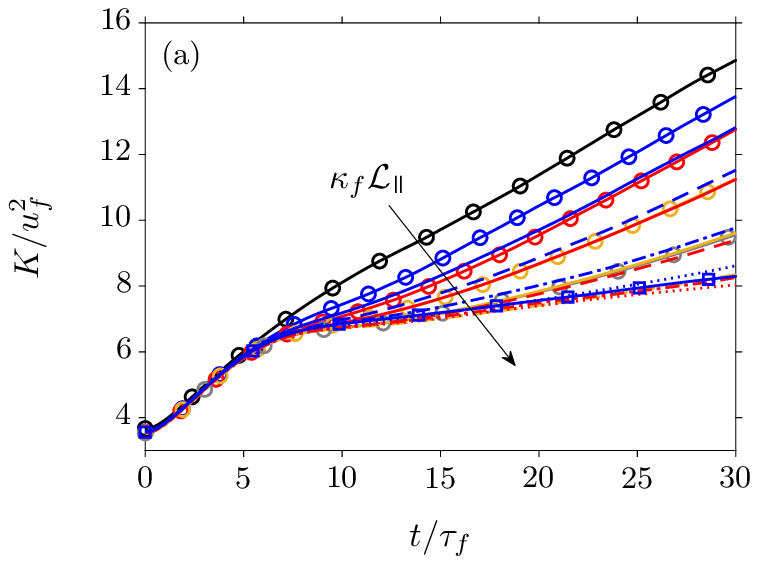}
          \label{fig:strong_kinetic_energy}}
    \hfill
  \subfloat
          {\includegraphics{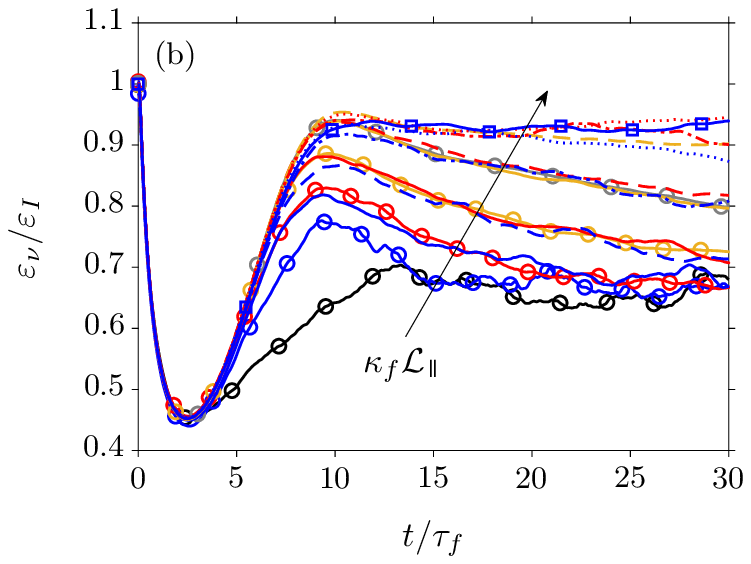}
          {\label{fig:strong_dissipation}}}
  \caption{
  Time evolution of box-averaged kinetic energy (a) and energy
  dissipation rate (b) for $\romanrossbyeps\approx0.06$ (strong rotation). Lines
  corresponding to same $\wndpp$ are grouped by color:
  $\wndpp=2$ (\filledblacksquare),
  $\wndpp=4$ (\filledbluesquare),
  $\wndpp=8$ (\filledredsquare),
  $\wndpp=16$ (\filledorangesquare),
  $\wndpp=32$ (\filledgreysquare).
  Lines corresponding to the same $A_r$ are grouped by line types:
  $A_r=1$ (\blackline\opencircle\blackline),
  $A_r=2$ (\blackline),
  $A_r=4$ (\blackdashedline),
  $A_r=8$ (\blackdashedtwoline),
  $A_r=16$ (\blackdashedthreeline),
  $A_r=32$ (\blackline\opensquare\blackline), cf. \cref{tb:perfosimul}.
  }
  \label{fig:strong}
\end{figure*}
\begin{figure*}[t]
  \centering
  \captionsetup[subfigure]{labelformat=empty}
  \subfloat
          {\includegraphics{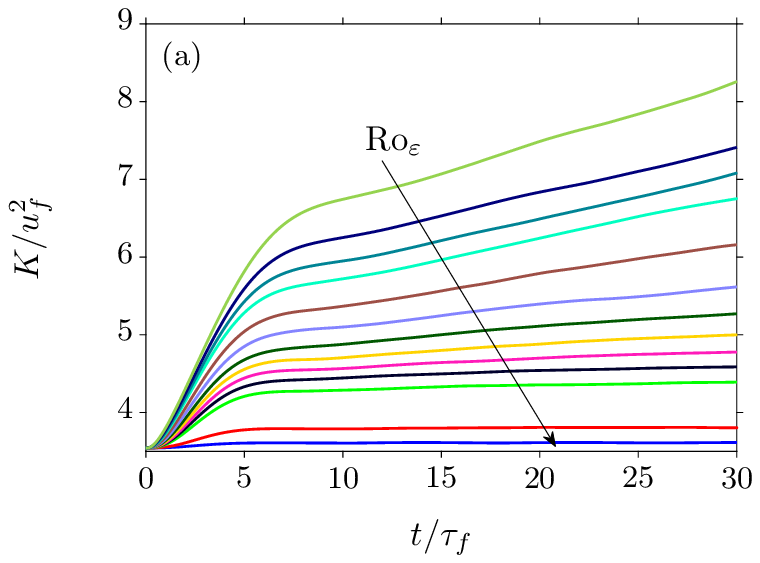}
          \label{fig:omgseries_kinetic_energy}}
    \hfill
  \subfloat
          {\includegraphics{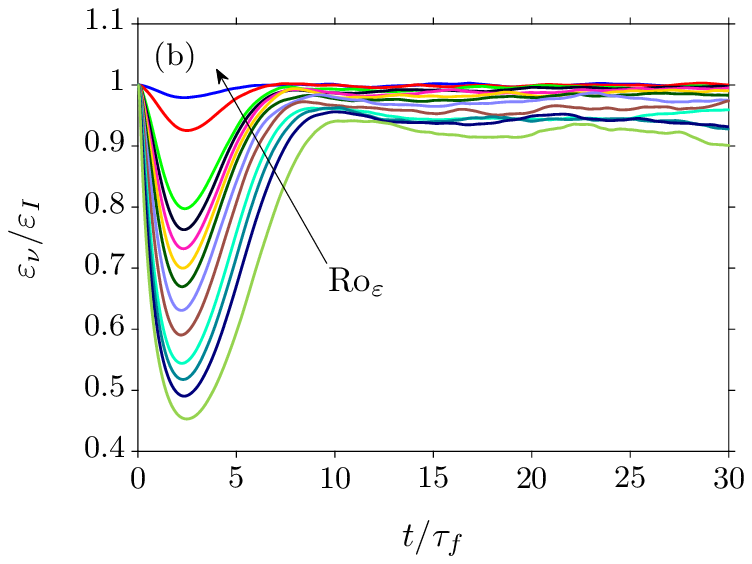}
          \label{fig:omgseries_dissipation}}
  \caption{Time evolution of box-averaged kinetic energy (a) and energy
  dissipation rate (b) for $\wndpp=8$ and $\wndpar=64$. Different line colors
  correspond to the range $0.06<\romanrossbyeps<1.25$, see
  \cref{tb:perfosimul}.}
  \label{fig:kf8e8_omgseries}
\end{figure*}
\begin{figure*}[t]
  \centering
  \captionsetup[subfigure]{labelformat=empty}
  \subfloat
          {\includegraphics{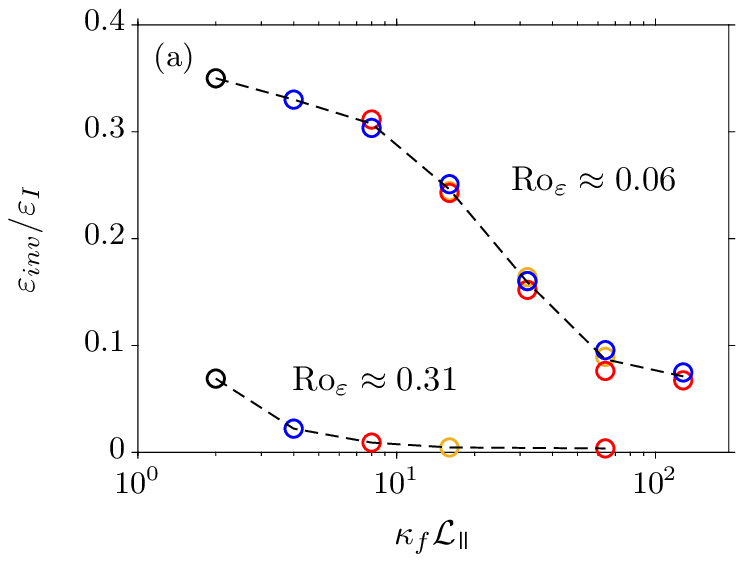}
          \label{fig:domain_invenergy}}
    \hfill
  \subfloat
          {\includegraphics{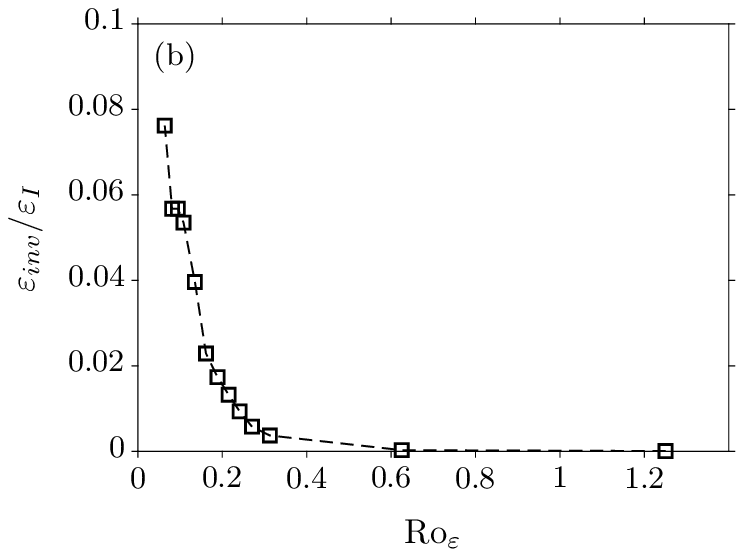}
          \label{fig:omgseries_invenergy}}
  \caption{Phase transition diagram  for weak and strong rotation and
  varying geometrical dimensions (a) and for constant geometrical dimension and
  varying $\romanrossbyeps$ (b). Color scheme of (a) is the same as in
  \cref{fig:strong}. In (a), the data point for $\wndpp=\wndpar=32$ (case
  \texttt{kf32-a01}) is almost identical to case \texttt{kf04-a08}
  ($\wndpp=4$; $\wndpar=32$), and is therefore not visible.}
  \label{fig:invenergy}
\end{figure*}

\section{Results}
First we assess the effects of geometrical dimension and rotation on the time
evolution of box-averaged kinetic energy $K$ and viscous dissipation
$\dissipnu$. The non-dimensional geometric parameters $\wndpp$ and $\wndpar$ are
varied for two fixed rotation rates: weak ($\romanrossbyeps=0.31$;
\cref{fig:weak}) and strong ($\romanrossbyeps=0.06$; \cref{fig:strong}).
Additionally, for a fixed and large domain, $\wndpp=8$ and $\wndpar=64$ (case
\texttt{kf08-a08}; \cref{fig:kf8e8_omgseries}), we investigate the Rossby number
range $0.06<\romanrossbyeps<1.25$. For more details about the simulation
parameters, please refer to \cref{tb:perfosimul}.

All cases undergo a transient of roughly $10\,\tf$ from the onset of rotation
(\cref{fig:weak,fig:strong,fig:kf8e8_omgseries}), which converges towards a
unique solution for sufficiently large $\wndpar$. We find that the results are
independent of the transversal domain size for $\wndpp\ge4$; see
\cref{fig:strong}, where the lines for different $\wndpp$ and identical
$\wndpar$ coincide. Departing from an isotropic state, where the energy cascade
is strictly forward ($\dissipnu/\dissipI=1$), $\dissipnu$ decreases
monotonically until it is lowest at approximately $3\,\tf$
(\cref{fig:weak_dissipation,fig:strong_dissipation,fig:omgseries_dissipation}).
For fixed $\romanrossbyeps$, \cref{fig:weak_dissipation,fig:strong_dissipation}
show that both $\wndpp$ and $\wndpar$ have no influence on the minimum of
$\dissipnu$. On the other hand, \cref{fig:omgseries_dissipation} suggests a
direct proportionality between the minimum value of $\dissip_{\nu}$ and
$\romanrossbyeps$.

\begin{figure}[t]
  \centering
  \includegraphics{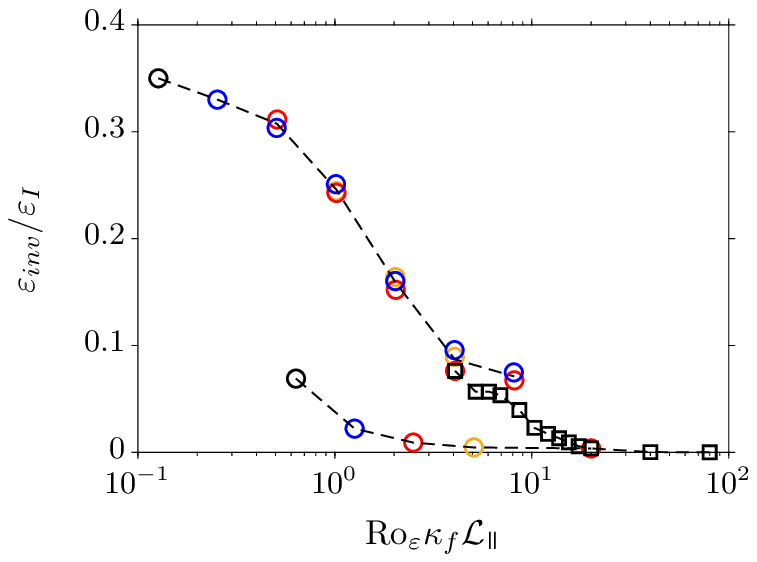}
  \caption{
  Phase transition diagram in terms of combined control parameter
  $\romanrossbyeps\wndpar$ for all data points of \cref{fig:invenergy}. Colored
  circles represent data from \cref{fig:domain_invenergy}, and squares data from
  \cref{fig:omgseries_invenergy}.}
  \label{fig:phasetransition}
\end{figure}

After $t\approx3\,\tf$, $\dissipnu$ increases towards $\dissipI$. Nevertheless,
the strong and weak rotation cases lead to a different final state for
$\dissipnu$. While increasing $\wndpar$ restores $\dissipnu=\dissipI$ for the
weak rotating case (\cref{fig:weak_dissipation}), the imbalance $\dissipnu <
\dissipI$, although lower than $0.075\,\dissipI$ for $\wndpar = 128$, persists
up to the final time for the strong rotating case
(\cref{fig:strong_dissipation}). Similarly to \cref{fig:weak_dissipation},
increasing $\romanrossbyeps$ reestablishes a forward energy cascade for a fixed
domain size (\cref{fig:omgseries_dissipation}). After the initial transient
($t>10\,\tf$), $\dissipnu$ follows mostly a slow linear decay
(\cref{fig:strong_dissipation}) or remains nearly constant
(\cref{fig:weak_dissipation,fig:omgseries_dissipation}). Consequently, $K$,
which evolves in time as $\dinline{K}{t}=\dissipI-\dissipnu$, grows
quasi-linearly
(\cref{fig:weak_kinetic_energy,fig:strong_kinetic_energy,fig:omgseries_kinetic_energy}).
Based on this idea we define the inverse energy flux $\dissipinv=\dissipI -
\dissipnu$ from the imbalance between energy injection rate and viscous
dissipation. To estimate $\dissipinv$, which is equal to the local slope of
$K(t)$, a linear least-square fit is applied to $15\,\tf<t<30\,\tf$ in
the time evolution of $K$
(\cref{fig:weak_kinetic_energy,fig:strong_kinetic_energy,fig:omgseries_kinetic_energy}).
The r.m.s. residual between the actual and fitted data indicates that
the linear regression model is appropriate. For the worst case,
\texttt{kf04-a08}, the r.m.s. residual is $0.65\%$ of the mean value. Assuming
that the linear law is exact and the noise is essentially Gaussian, one obtains
$0.0004$ for the standard error of the slope coefficient.
Results for the inverse energy flux are thus shown in
\cref{fig:invenergy,fig:phasetransition} in form of a phase transition diagram.

From \cref{fig:domain_invenergy}, we see that the inverse energy flux
$\dissipinv$ decreases monotonically with $\wndpar$ for both
$\romanrossbyeps\approx0.31$ and $\romanrossbyeps\approx0.06$. Moreover, results
for the strong rotating case suggest that increasing $\wndpp$ while retaining
$\wndpar$ leads to negligible differences in $\dissipinv$~\textemdash~see the
overlapping circles with different colors for $\romanrossbyeps\approx0.06$.
Transition from a split to a forward cascade system occurs gradually. For
$\romanrossbyeps\approx0.31$ and $\wndpar=64$ less than $0.004\,\dissipI$ is
transferred in the inverse direction, whereas for $\romanrossbyeps\approx0.06$ a
split cascade is still present at $\wndpar=128$. For a fixed domain size with
$\wndpp=8$ and $\wndpar=64$ (case \texttt{kf08-a08};
\cref{fig:omgseries_invenergy}), $\dissipinv$ is continuously suppressed for
increasing $\romanrossbyeps$ and transition to a forward cascade system occurs
in the vicinity of $\romanrossbyeps=1$.

A question that follows from these results is for which combination of governing
non-dimensional parameters regime transition occurs. From literature, a possible
criteria is $\romanrossbyeps\wndpar = C$, where $C$ is a constant
\citep{Seshasayanan2018,Alexakis2018}. To test this hypothesis,
\cref{fig:phasetransition} presents the data from \cref{fig:invenergy}, but
juxtaposed in a single diagram and scaled accordingly with
$\romanrossbyeps\wndpar$. The curves for different $\romanrossbyeps$ do not line
up; hence, this criteria disagrees with our data. A discussion on a possible
reason is given in the next section.

\begin{figure}[t]
  \centering
  \captionsetup[subfigure]{labelformat=empty}
  \subfloat
          {\includegraphics{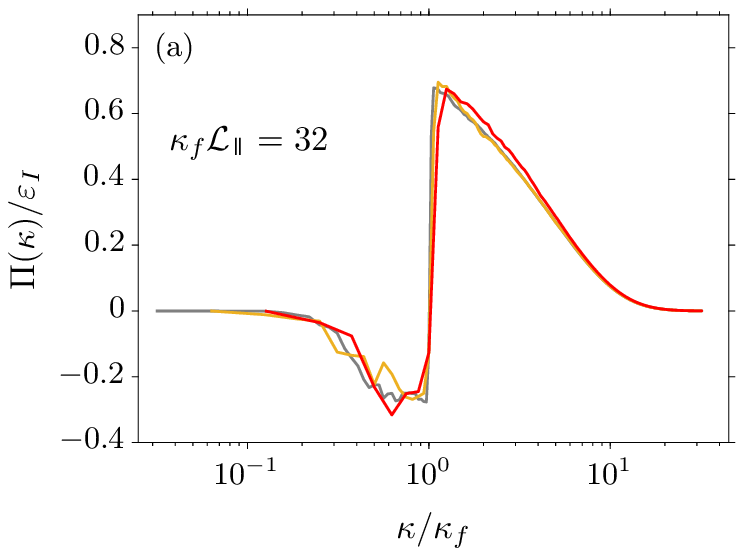}
          \label{fig:specfx_const_kflpar}}
  \\[-1ex]
  \subfloat
          {\includegraphics{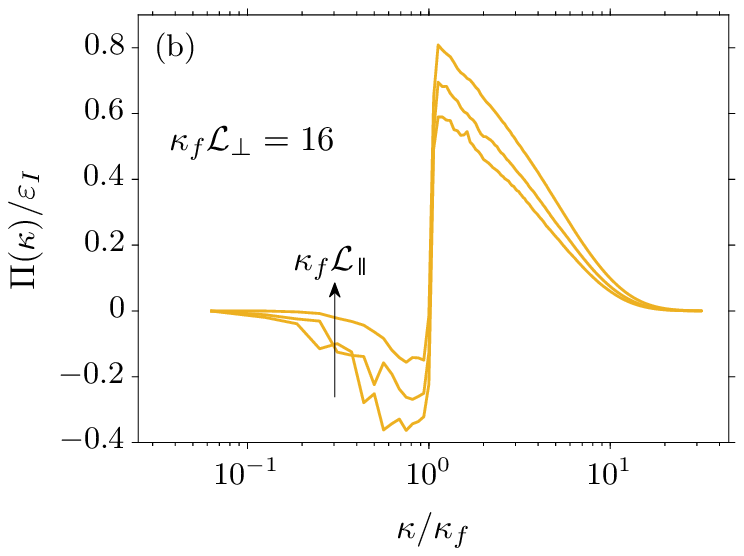}
          \label{fig:specfx_const_kflpp}}
  \caption{Spectral energy flux for $\romanrossbyeps\approx0.06$ and cases with
  $\wndpar=32$ (a) and $\wndpp=16$ (b). In (a), $\wndpp=8$ (\redline),
  $\wndpp=16$ (\orangeline) and $\wndpp=32$ (\greyline). In (b), $\wndpar=16,32$
  and $64$ (\orangeline). Arrow denotes the direction of increase.}
  \label{fig:specfx}
\end{figure}

Now we turn our attention to the influence of $\wndpar$ and $\wndpp$ on the
spectral energy flux and energy spectra. Hereafter we present results for the
strong rotating case with $\romanrossbyeps\approx0.06$ only, as differences are
more pronounced than in the weak rotating case. Although we show instantaneous
data at $t=30\,\tf$, the trend described in what follows also holds for other
instants of time. Conservation of energy requires the portion of the injected
energy that is not dissipated to be accumulated. By analyzing the spectral
energy flux $\Pi(\wn)$, we find that the net energy transfer
$T(\wn)=-\dinline{\Pi}{\wn}$ is positive for $\wn<\wnf$. In other words,
wavenumbers in this range gain energy and we observe an upscale energy transfer.
Evidence is presented in \cref{fig:specfx}, which also highlights how sensitive
$\Pi(\wn)$ is with respect to changes in $\wndpar$ and $\wndpp$. In this regard,
\cref{fig:specfx_const_kflpar}, where $\wndpar$ is constant and $\wndpp=\{8$,
$16$, $32\}$, shows that the shape of $\Pi(\wn)$ remains unaltered for different
$\wndpp$. On the other hand, varying $\wndpar$ from $16$ to $64$ while $\wndpp$
is constant, reduces the magnitude of the inverse energy flux and the range of
wavenumbers for which an upscale energy transfer takes place, see
\cref{fig:specfx_const_kflpp}. Therein, greater values of $\wndpar$ are also
associated with an enhanced spectral energy flux for $\wn >
\wnf$. This is a consequence of the fixed energy input rate $\dissipI$, which
causes the step in $\Pi(\wn)$ at $\wn=\wnf$ to be the same for all cases.

\begin{figure}[t]
  \centering
  \captionsetup[subfigure]{labelformat=empty}
  \subfloat
          {\includegraphics{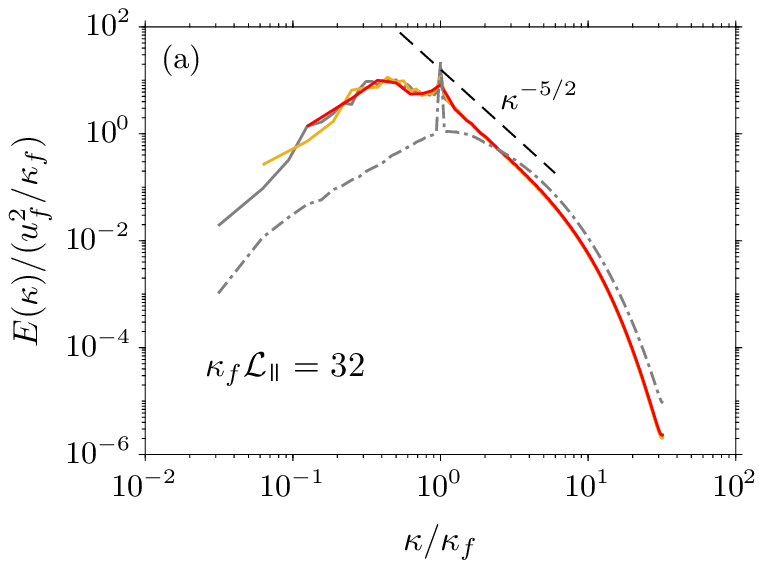}
          \label{fig:strong_3denergyspectra_kflpar}}
  \\[-1ex]
  \subfloat
          {\includegraphics{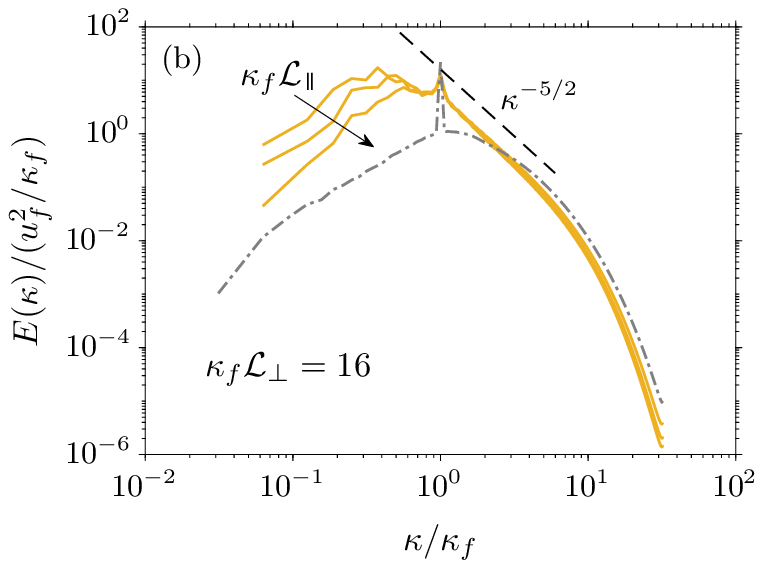}
          \label{fig:strong_3denergyspectra_kflpp}}
  \caption{Three-dimensional spherically averaged energy spectrum for
  $\wndpar=32$ (a) and $\wndpp=16$ (b) with $\romanrossbyeps\approx0.06$. Line
  styles are the same as in \cref{fig:specfx}, apart from the reference energy
  spectrum of \cref{fig:iso_energyspec} with $\wndpp=\wndpar=32$
  (\greydashedtwoline).}
  \label{fig:strong_3denergyspectra}
\end{figure}

The three-dimensional energy spectra $E(\wn)$ for the same cases are shown in
\cref{fig:strong_3denergyspectra}.  Additionally, the energy spectrum of case
\texttt{kf32-a01} with $\wndpar=\wndpp=32$ from \cref{fig:iso_energyspec} at the
onset of rotation is included as reference.
\Cref{fig:strong_3denergyspectra_kflpar} reinforces that $\wndpar$ dictates the
degree of energy accumulation, as the curves for different $\wndpp$ and constant
$\wndpar$ overlap. In agreement with results in \cref{fig:specfx} for
$\Pi(\wn)$, we observe significantly higher levels of energy for $\wn<\wnf$ with
respect to the isotropic reference spectrum. These are reduced for increasing
$\wndpar$, see \cref{fig:strong_3denergyspectra_kflpp}.


\begin{figure}[t]
  \centering
  \includegraphics{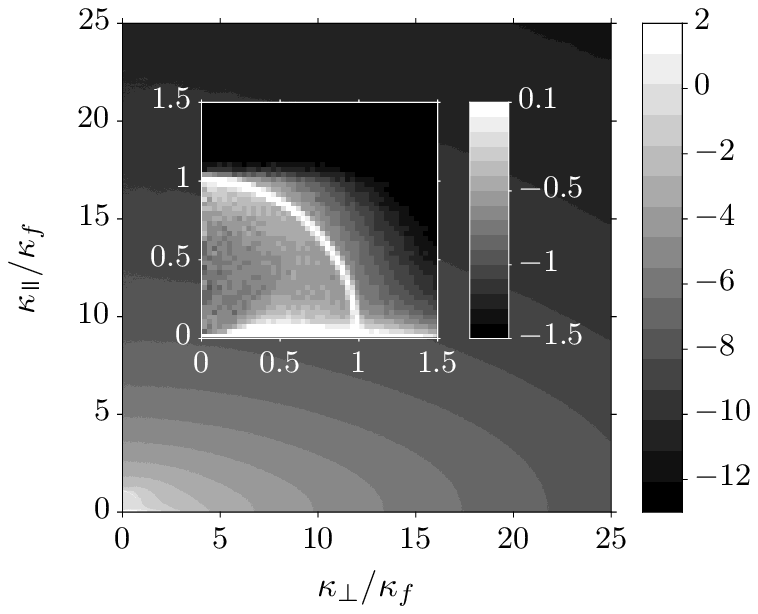}
  \caption{
  Two-dimensional energy spectrum for $\romanrossbyeps\approx0.06$ with
  $\wndpp=\wndpar=32$ (case \texttt{kf32-a01}) at $t=30\,\tf$. Data is
  normalized by $(2\pi\wnpp)\,\uf^2/\wnf^3$ and plotted in $\log_{10}$. The
  inset highlights the region around the forcing wavenumber: $\wnpp/\wnf<1.5$
  and $\wnpar/\wnf<1.5$.}
  \label{fig:ecyl_kf32}
\end{figure}

As for the distribution of energy in terms of $\wnpar$ and $\wnpp$,
\cref{fig:ecyl_kf32} presents the two-dimensional energy spectrum
$E(\wnpp,\wnpar)$. Results are shown exclusively for case \texttt{kf32-a01} with
$\wndpp=\wndpar=32$, as it contains most large scale resolution. The energy
spectrum is non-dimensionalized with $2\pi\wnpp$, in such a way that contour
levels of isotropic spectra appear as circles centered at the origin. In
agreement with previous works, \cref{fig:ecyl_kf32} confirms that the kinetic
energy has the tendency to accumulate at lower $\wnpar/\wnf$. Hence,
$E(\wnpp,\wnpar)$ is anisotropic and contour levels display an elliptical shape
with major axis aligned with the $\wnpp\text{-direction}$. This is
observed even for high wavenumbers and suggests that all scales of motion are
influenced by rotation; indeed, for this case, $\kzeman\lkolmo = 1.1$, where
$\kzeman=(\Omega^3/\dissipI)^{1/2}$ is the Zeman wavenumber \citep{Delache2014}.
At the same time, the energy input remains isotropic. See the inset for the
imprint of the isotropic forcing scheme, which delineates the bright area
located at $\wnpar^2+\wnpp^2=\wnf^2$. In addition, we see higher energy levels
in the vicinity of $\wnpar/\wnf=0$.

\begin{figure}[t]
  \centering
  \captionsetup[subfigure]{labelformat=empty}
  \subfloat
          {\includegraphics{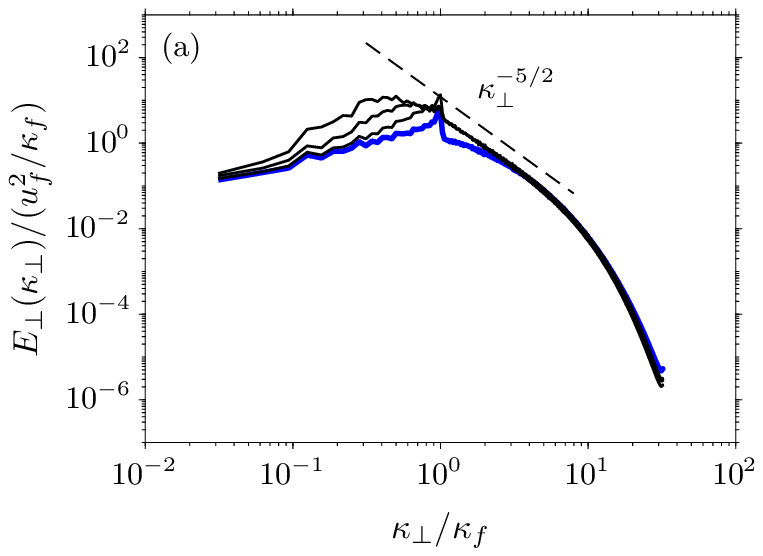}
          \label{fig:epp_kf32}}
  \\[-1ex]
  \subfloat
          {\includegraphics{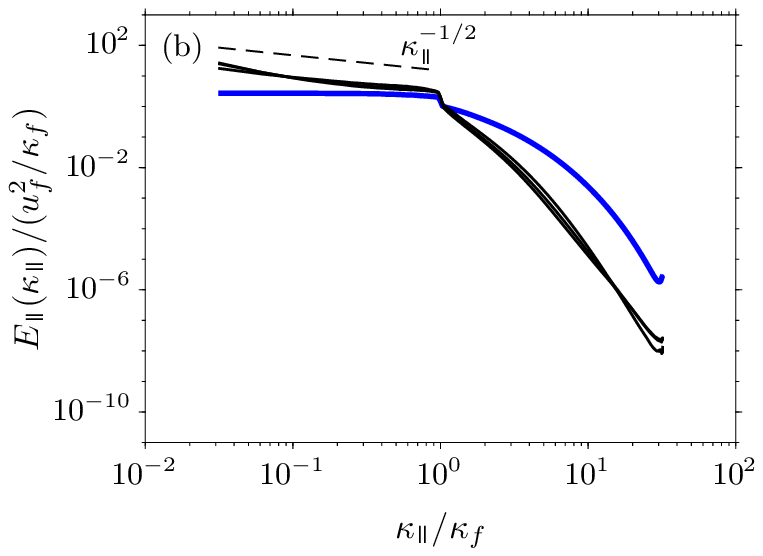}
          \label{fig:epar_kf32}}
  \caption{
  One-dimensional energy spectra for $\romanrossbyeps\approx0.06$ and
  $\wndpp=\wndpar=32$ (case \texttt{kf32-a01}) along directions $\wnpp$ (a) and
  $\wnpar$ (b). Lines represent the time evolution of the energy spectrum: $t=0$
  (\blueline), and $t=10, 20$ and $30\,\tf$ (\blackline). A reference line for
  the scaling laws that best agrees with the presented data is also shown
  (\blackdashedline). }
  \label{fig:epp_epar_kf32}
\end{figure}

An anisotropic distribution of energy is predicted by the weak inertial-wave
theory, which suggests that the energy spectrum has the form $E(\wnpp,\wnpar)
\sim \wnpp^{-5/2} \wnpar^{-1/2}$~\citep{Galtier2003}. To test if our data
presents any sign of this scaling law, we show in \cref{fig:epp_epar_kf32}
instantaneous one-dimensional energy spectra along the perpendicular and
parallel directions, i.e. $E_\perp(\wnpp)$ and $E_\mypara(\wnpar)$ for
$t=0,10,20$ and $30\,\tf$. \Cref{fig:epp_kf32} shows that energy levels increase
progressively for $\wnpp <
\wnf$, whereas for $\wnpp > \wnf$, the distribution of energy is nearly
unaltered. Also for $\wnpp>\wnf$, we observe that a narrow wavenumber range
develops from the initial state and approaches best a $\wnpp^{-5/2}$ scaling
law. Regarding $E_\mypara(\wnpar)$, \cref{fig:epar_kf32}, the energy content for
$\wnpar>\wnf$ is significantly lower than at the onset of rotation. This
corroborates the idea that rotation lessen the flow field dependency on the
direction parallel to the rotation axis. As time evolves, the range
$\wnpar<\wnf$ resembles best a $\wnpar^{-1/2}$ scaling law for all time
instants. We emphasize that this result is
essentially different from predictions of the weak inertial-wave theory, as the
latter estimates $E(\wnpar)\sim\wnpar^{-1/2}$ for $\wnpar$ larger than the
forcing wavenumber.

\section{Discussion}

This work investigated through direct numerical simulations the effects
of domain size and rotation rate on the energy cascade direction of rotating
turbulence. The data here presented add substantially to previous works, which,
in contrast, focused on smaller and shallower domains ($\wndpar$ and $\wndpp$
$<8$ \citep{Smith1996a,Deusebio2014}). The presented results, therefore,
contribute towards a complete picture of the phase diagram, which unveils
transition from inverse to forward through a split energy cascade in rotating
turbulence.

Our results support $\wndpar$ as the primary control parameter provided that
$\romanrossbyeps$ is constant and $\wndpp>4$. In this scenario, transversal
finite-size effects of $\wndpp$ on the inverse energy transfer $\dissipinv$ are
negligible for our cases with aspect ratio $A_r \geq 1$. For weak rotation with
$\romanrossbyeps\approx0.31$, transition from a split to a forward cascade was
observed at $\wndpar\approx64$. For the strong rotating case, however, although
strongly suppressed, a portion of the injected energy ($\dissipinv \approx
0.075\,\dissipI$) still cascaded inversely and accumulated at the large scales
for $\wndpar=128$.

We attribute the fact that $\dissipinv$ does not become exactly zero for
$\romanrossbyeps\approx0.31$ to two effects. First, the simulations considered
in this study are limited to $\romanreynoldslmbda\approx 68$. A higher
Reynolds number could contribute to a stronger forward cascade, possibly
reducing $\dissipinv$ to zero. Second, although effects of the geometric
non-dimensional parameter $\wndpp$ are minor, results hint that larger values of
$\wndpp$ could also contribute to a reduction of $\dissipinv$. In this manner,
indefinite increase of $\wndpp$ could potentially change the phase diagram in
the vicinity of $\dissipinv/\dissipI=0$, and could cause regime transition to be
sharp rather than smooth. The recent study of \citet{Benavides2017} has shown
that a continuous increase of horizontal domain dimensions shifts the transition
behavior for thin layer turbulence from smooth to critical. We hope that further
studies will help to fill the parameter space for higher Reynolds numbers and
even longer domain sizes.

For $\romanrossbyeps\approx0.06$, we agree with \citet{Deusebio2014} and
believe that a continuous increase of $\wndpar$ would result in transition to a
forward energy cascade. Nevertheless, results for the weak case suggest a
slow-paced transition and significantly larger values for $\wndpar$ might be
required. Interestingly, the transition of $\dissipinv$ in terms of $\wndpar$
resembles a logistic function, similar to what has been
found for regime transitions in thin layer turbulence
\citep{Benavides2017}.

In search of a criteria for transition between a forward and a split cascade
system, we made an attempt to express $\dissipinv/\dissipI$ for all parameter
points as a function of $\romanrossbyeps \wndpar$. As the different curves do
not overlap, we believe that a criteria for transition should stem from a more
general match of timescales. A criteria such as $\romanrossbyeps\wndpar = C$,
can be obtained by requiring the slowest inertial wave frequency
$1/\twave=2\Omega/\wndpar$ and the eddy turnover frequency $\uf\wnf$ at the
forcing scale to be of same order \citep{Alexakis2018,Seshasayanan2018}.
Alternatively, we can frame the problem within the idea that rotation alters the
spectral transfer time $\tau_s$ at which energy is transferred to smaller
scales. Thus, it follows that $\dissipnu
\sim \uell^2 / \tau_s$, with $\uell$ a velocity scale characteristic of
eddies of size $\ell$, and $\tau_s \sim \tau_{nl}^2/\tau_3$
\citep{Kraichnan1965,zhou1995,Galtier2003}. Here, $\tau_{nl}\sim\ell/\uell$ is
the nonlinear timescale and $\tau_3$ is the relaxation time of triple velocity
correlations. The relaxation time in isotropic turbulence simplifies to
$\tau_{nl}$ to recover the dissipation law, i.e. $\dissipnu \sim \uell^3 /
\ell$.

Now the condition $\romanrossbyeps \wndpar=C$ can be obtained by requiring
$\dissipnu=\dissipI$, and assuming $u \sim \uf$, $\tau_{nl} \sim \tf$ and
$\tau_3\sim \twave$. So, $\romanrossbyeps \wndpar=C$ is equivalent to state that
in the presence of rotation the nonlinear timescale remains of the order of
$\tf$, and that the relaxation timescale $\tau_3$ is given by the inverse of the
slowest inertial wave frequency, i.e $\tau_3\sim\twave$. A generalization of the
previous reasoning would be to consider a $\tau_{nl}$ obtained from a measured
velocity quantity, like the r.m.s velocity, and the lengthscale $\ell$ possibly
as $\ell_\perp$, as the triadic interactions are expected to be depleted in the
direction parallel to the rotation axis \citep{Nazarenko2011b}. The relaxation
time $\tau_3$ could be sought as a function of both $\tf$ and $\tomg$. In this
manner, more general criteria like $\romanrossbyeps^a (\wndpar)^b=C$ arise,
where $a$ and $b$ are yet undetermined exponents.

Results for scaling laws of the energy spectrum are here not conclusive, and
there is no clear sign of an inertial range over several decades. This is
plausible since our initial and isotropic field with $\romanreynoldslmbda
\approx 68$ does not contain a clear inertial range. In spite of that, the
narrow wavenumber region after $\wnpp = \wnf$ develops and approaches best a
$\wnpp^{-5/2}$ scaling law. Our results also show that, the $\wnpp^{-5/2}$ and
$\wnpar^{-1/2}$ scalings appear at different wavenumber ranges, and that the
$\wn^{-5/2}$ scaling prevails in the 3D energy spectrum, see
\cref{fig:strong_3denergyspectra}.

 
\bibliography{manuscript}
\end{document}